# Puffy: A Step-by-step Guide to Craft Bio-inspired Artifacts with Interactive Materiality

**Sark Pangrui Xing**[1,2]
sark.xing@connect.polyu.hk

**Bart van Dijk**[2]
contact@b-vandijk.nl

**Pengcheng An**[3]
anpc@sustech.edu.cn

**Miguel Bruns**[2]
mbruns@tue.nl

**Yaliang Chuang**[2]
y.chuang@tue.nl

**Stephen Jia Wang**[1]
stephen.j.wang@polyu.edu.hk

School of Design
[1] The Hong Kong Polytechnic University
Hong Kong SAR, China

Industrial Design
[2] Eindhoven University of Technology
Eindhoven, The Netherlands

School of Design
[3] Southern University of Science and Technology
Shenzhen, China

**Abstract**

A rising number of HCI scholars have begun to use materiality as a starting point for exploring the design's potential and restrictions. Despite the theoretical flourishing, the practical design process and instruction for beginner practitioners are still in scarcity. We leveraged the pictorial format to illustrate our crafting process of Puffy, a bio-inspired artifact that features a cilia-mimetic surface expressing anthropomorphic qualities through shape changes. Our approach consists of three key activities (i.e., *analysis*, *synthesis*, and *detailing*) interlaced recursively throughout the journey. Using this approach, we analyzed different input sources, synthesized peers' critiques and self-reflection, and detailed the designed experience with iterative prototypes. Building on a reflective analysis of our approach, we concluded with a set of practical implications and design recommendations to inform other practitioners to initiate their investigations in interactive materiality.

**Introduction**

The field of HCI has developed a growing interest (e.g., [4],[11],[15],[27]) in addressing materiality as an entry for design research [10] by entangling miniaturized sensing and actuating technologies with advanced materials [19]. Such entanglement has enabled HCI scholars to create matters with shape-changing [1,26] quality, interactive materiality [30], expressivity [2], to name but a few, navigating HCI research toward its next wave [7]. In response, explorations in forms of design notions (e.g., materiality of information[5], materiality of interaction [38], form-giving [36]), design methods (e.g., [10,16,39]) have sparked among the HCI communities.

While researchers are still developing theories to support this emerging field, many have developed tools to support concept ideation e.g., [45] and prototyping (e.g., [20,22,42]). In particular, Morphino [24] has proposed a card-based kit that archives illustrations of how organisms change shape, allowing practitioners to widely explore suitable shape-changing mechanisms from nature. Additionally, design practitioners have explored adopting materiality in practices for behavior change [17,29,33], lived experiences [43,44], emotional expression [13], smart garment [34,41] to name but few. Despite those prior works, it seems few studies address on the process and rationale of how their shape-changing concepts were ideated and concretized; and how the materiality is mapped and correlated with user interactions. In other words, the overarching question is how are certain qualities of organic shape changes, such as subtlety, ambiguity, and temporality transformed into the materiality of shape-changing?

To address this question, we present a mixed materiality approach by which we yielded Puffy, a shape-changing artifact that spontaneously changes its shape and materiality to express its emotions. Our design process consisting of three key activities (A-S-D, i.e., *analysis*, *synthesis*, and *detailing*, see p. 2) is mainly based on two approaches, namely biomimicry [25] and interactive materiality [30]. We used biomimicry-alike to generate implications serving as a source of inspiration. In the follow-up steps, we analyzed different sources of input gained in various stages, synthesized self-reflections and peers' critiques, and detailed the design through iterative prototyping.

In short, we contribute the HCI community a concrete interactive materiality design case and moreover we offer an illustrative step-by-step guide and prototyping suggestion to assist interaction design practitioners to design interactions that feature in rich temporality, interactive materiality, and computational aesthetics.

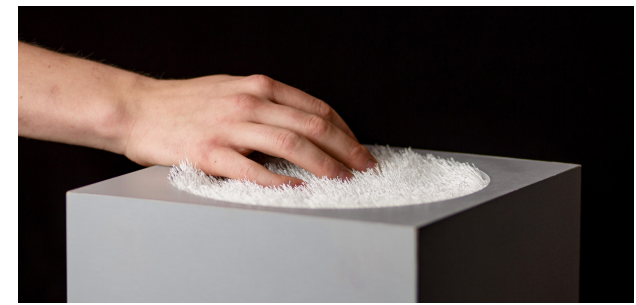

# Step-by-step Guide to Craft Bio-inspired Artifacts with Interactive Materiality

**ANALYSIS** *Analyzing various input* → **SYNTHESIS** *Synthesizing the findings* → **DETAILING** *Detailing the design outcome*

**1 Shape transitions**

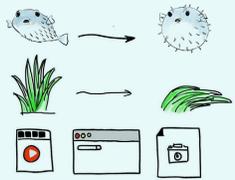

**2 Analogue materials**

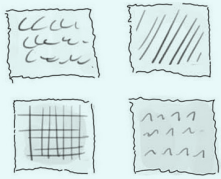

**3 Materials samples**

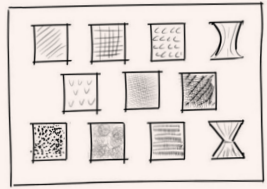

**4 Transition samples**

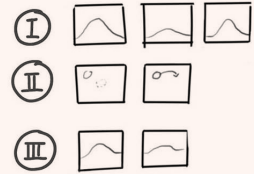

**5 Quick-and-dirty mimicking**

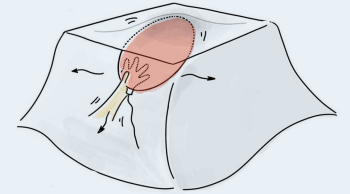

**8 Computational mechanism explorations**

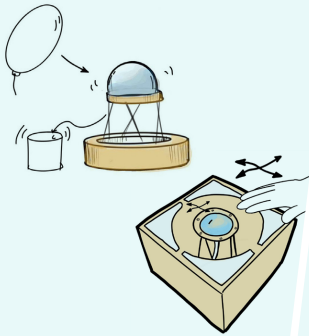

**9 Affirming the computational whole**

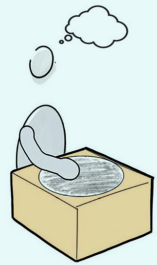

**7 Materiality experiencing setup**

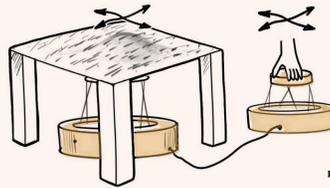

**6 Interaction mapping**

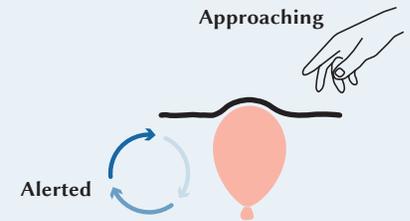

**10 Design critique**

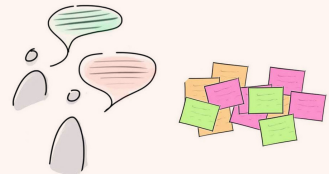

**11 Analyzing feedback**

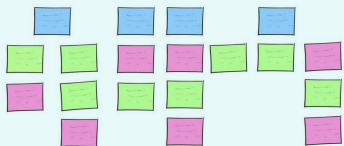

**12 Synthesizing feedback**

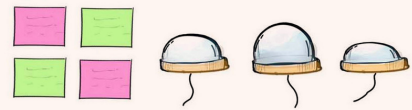

**13 Detailing the design artifact**

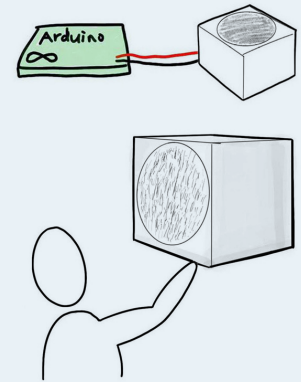

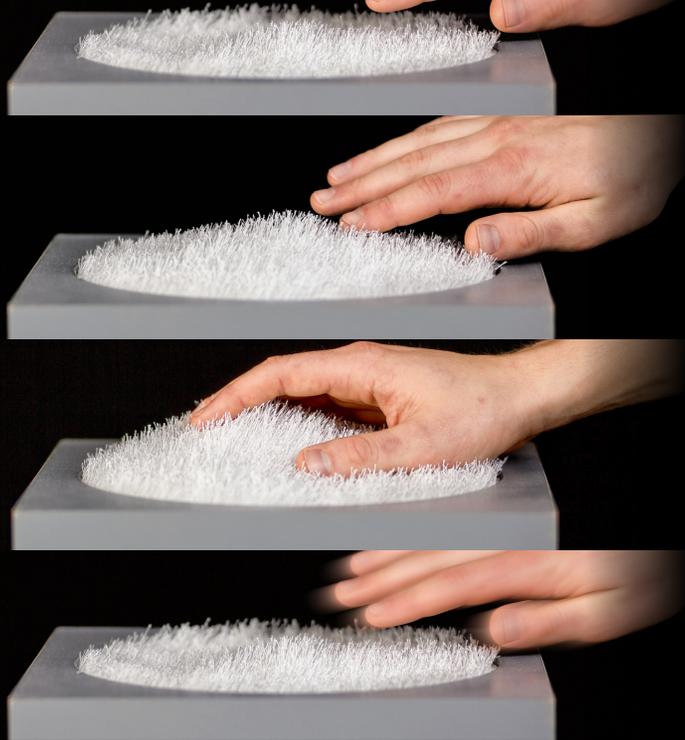

**The Puffy artifact**

Puffy is a cilia-mimetic interactive artifact that responds to human proximity and expresses "emotions" through shape changes. It mimics how a pufferfish responds to predators by straightening spikes and inflation. Depending on external surrounding such as a danger nearby, Puffy reveals three kinds of anthropomorphic characteristics (i.e., *Calm, Alerted, and Relaxed*) through the changes of haptic shapes and fabric sturdiness.

The conceptualization of Puffy followed the mimicking strategy described in [25] and the nature-inspired card deck [24]. Specifically, we looked at a number of lively inspirations that contain dynamic behavior from nature. However, instead of focusing on how *biologically* or *mechanically* a creature changes its shape, we, as interaction designers, focused on how the shape changes affect the aesthetic qualities (e.g., figures shown on p.3) of the creature, which influences how the creature is perceived and feelings that are evoked by the viewer. In other words, we were not simply mimicking the shape-changing mechanism, e.g., the puffer fish's accordion-like folded stomach, but rather the *experiences* evoked by the movement and *somaesthetics* derived from the selected natural element. We chose the most intriguing one – Pufferfish as our inspirational reference because of its spiky texture, rebelling behavior, and the correlations of these two with its surroundings. The subsequent sections will elaborate on how we transformed those pufferfish's characteristics into Puffy's interactive materiality.

# Analyzing various input *(ANALYSIS)*

## 1 Shape transitions

We first explored aesthetic inspiration from nature such as animals and plants. We found that the shape transitions of pufferfishes had some attractive attributes:

1. the shape change communicates *tension* as the fish expands or squeezes;

2. the growth and angle change of its spikes emphasizes its *repelling* emotion;

3. the underlying relationship between these changes and the pufferfish's *intention* and surroundings (e.g., the *predators*).

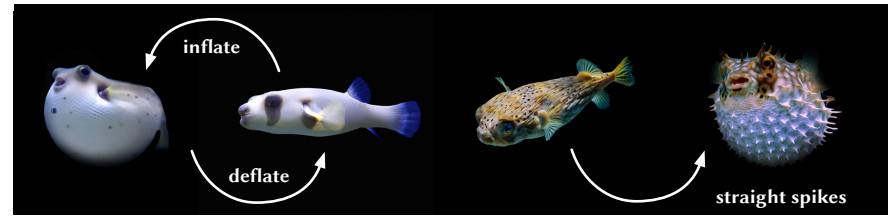

(Credit: Author, Author, Stelio Puccinelli, Martin Kleppe)
Inspiration sources: *Puffer Fish Puffing up when caught*, 2011. https://youtu.be/ccsvJMkF5Bs;
*Blow me, Beautiful*, 2013. https://youtu.be/S7y4quhmMW0;
*Dogface Puffer Fish Puffs Up Like Balloon*, 2019. https://youtu.be/-qf5vPq_z7U)

## 2 Analogue materials

Consecutively, to eliminate the confounding bias from the colors of the material, we purposely explored and examined extensive analogue materials that were in white and matched the interesting attributes of shape transitions gathered from the prior analysis. We obtained a profound and embodied understanding of the selected behavior through *first-hand* explorations, emphasizing the *visual* and *sensory* feelings.

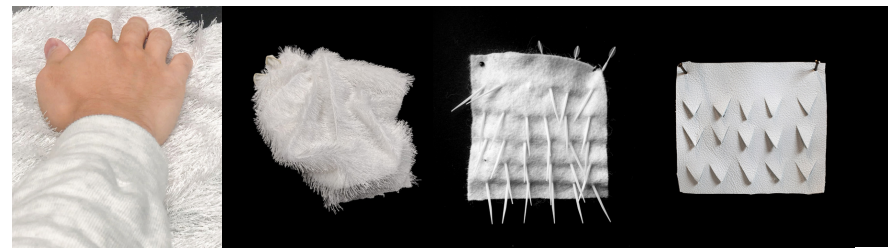

## 3 Material samples *(SYNTHESIS)*

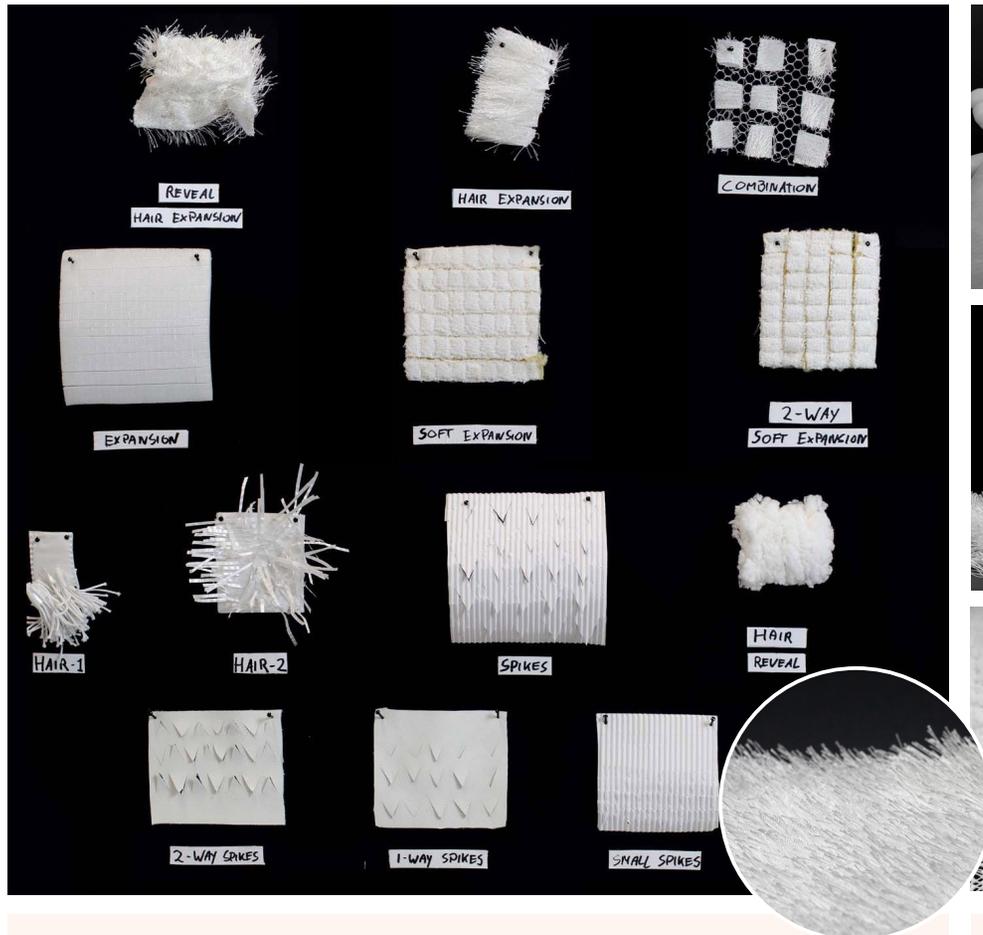

Among the explorations, some focused on manipulating material qualities, while some consisted of combining multiple materials to allow for expansion. Besides, we also looked at how we could produce our own materials with custom specifications (e.g., 3D printed hairy material, Cilllia [23]). While the exploration provided directions for material adaptation, we found that our transition could be best expressed through an un-adapted, hairy fabric combined with a shape-changing mechanism that worked best as it emphasizes the sturdy visual expression and haptic experience.

## 4 Transition samples *(SYNTHESIS)*

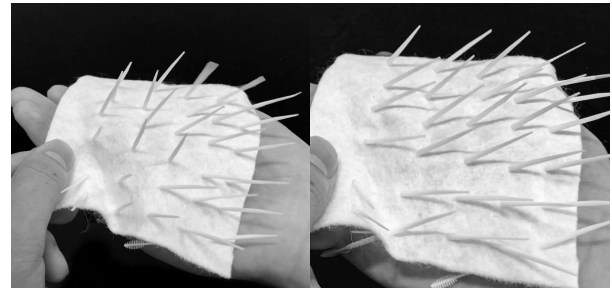 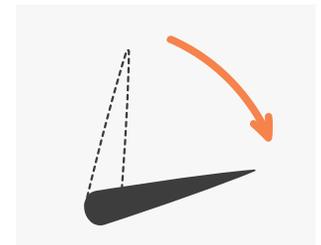

**(Straight** spikes to **flat)**

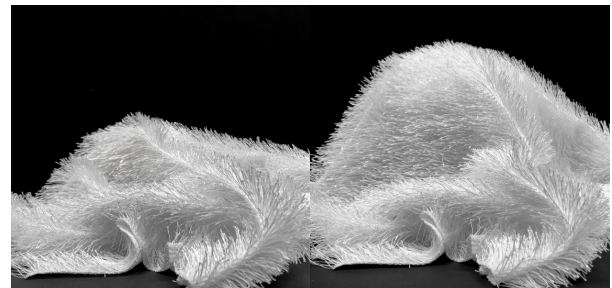 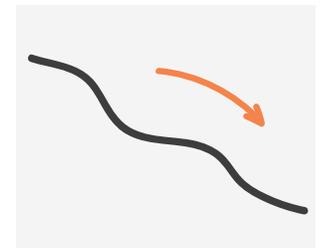

**(Non-linear** deflation)

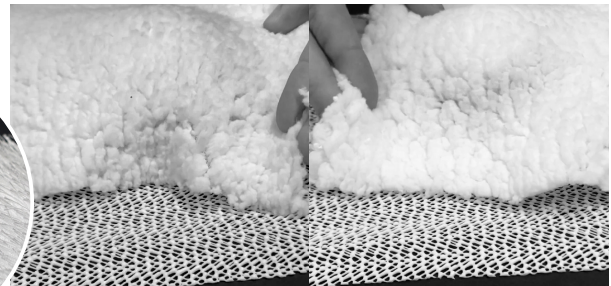 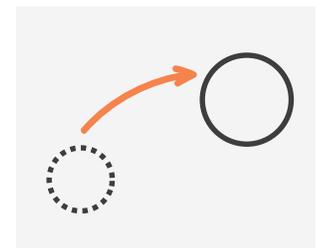

**(Movement** and **expansion)**

We used pufferfish as an inspiration to explore different transition techniques and aimed to find inherently coupled materials and transitions. We then synthesized our explorations regarding surface texture, shape transitions, and behavior movements respectively. To highlight the subtlety and temporality of the organic shape changes, we provided considerable attention to extracting the underlying information. For example, without a looping *replay* and multiple *rewinds* of the pufferfish video clips, we would not have been able to see that the deflation was gradual, sluggish, and *non-linear*, while the inflation was *abrupt* and *linear*.

# How will the *materiality* react to the user with the *synthesized materials* and *transitions*?

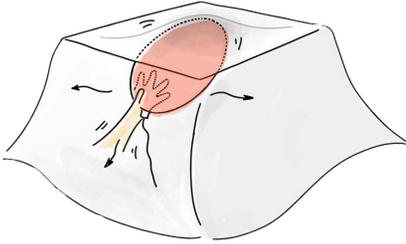

## 5 Quick & Dirty mimicking *(DETAILING)*

To answer the question from an experience perspective, we used a quick-and-dirty setup consisting of a texture of interest and a balloon placed underneath it. We mimicked the previously synthesized materials and transitions e.g., the spikes, the shape changes, and the underlying pufferfish-predator relationship. Through the mimicking, we were intrigued by how those reciprocally incorporated elements could be mapped and manifested in detailing and mimicking the desired materiality, resulting in an intriguing interaction design.

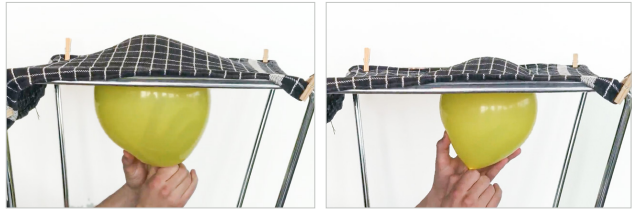

**Hand-controlled actuation**

As we intended to mimic the escaping behavior of pufferfish, we explored techniques that allow for rapid examination of materiality experiences, one of which inspired us the most is ClothSurface [12]. We consecutively replicated it through a quick-and-dirty setup in which a piece of fabric was wrapped around the legs of an upside-down chair. By manipulating a balloon placed underneath (i.e., inflating, deflating, or moving), the shape transitions could be mimicked, allowing us to experience and evaluate the aesthetic qualities promptly and haptically. At last, we were impressed by the hairy quality of the textile and the expansion of movement on the textile.

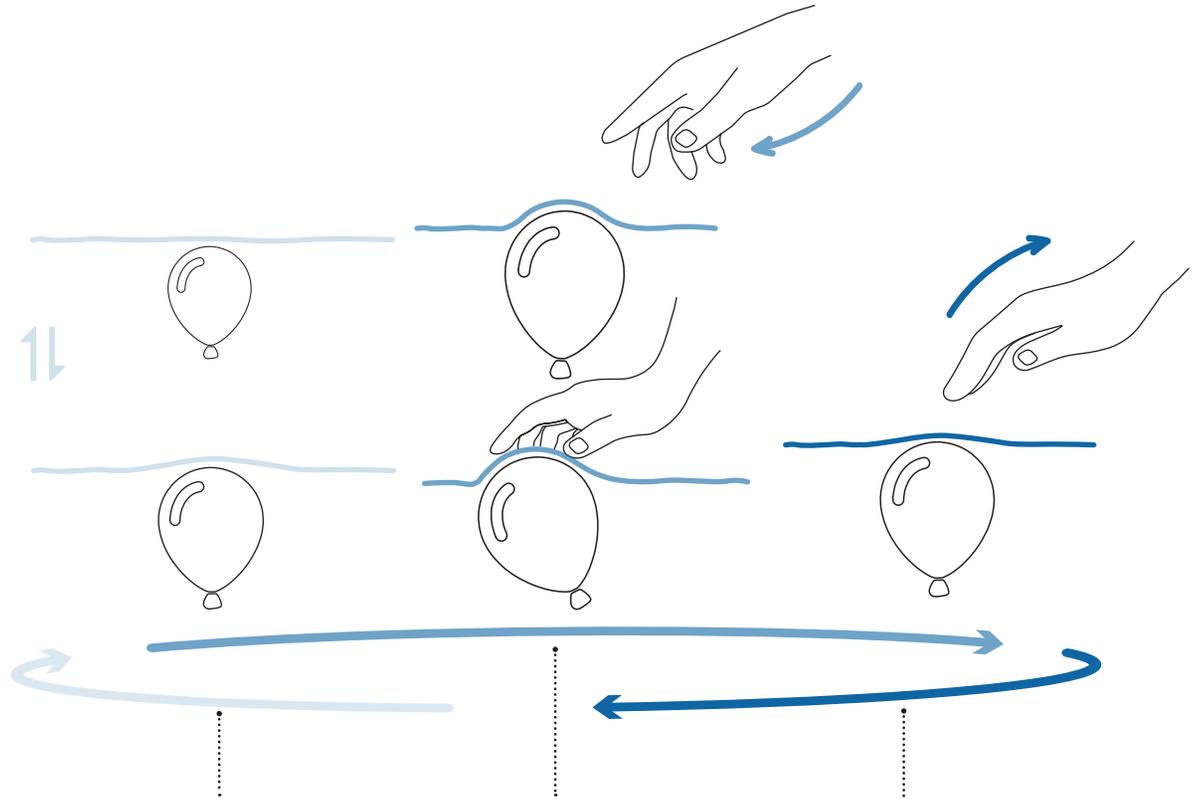

## 6 Interaction mapping *(DETAILING)*

| No interaction | Hand approaching | Hand leaving |
|---|---|---|
| **Calm** | **Alerted** | **Relaxed** |
| When no human interaction is involved, no significant changes in materiality will occur. The object stays calmly in back-and-forth looping transitions, behaving like 'breathing'. | When a user attempts to approach, the object becomes alerted and expands significantly. Its surface becomes more sturdy and consecutively moves always to escape from the 'danger'. | When the 'danger' disappears, the object becomes relaxed and gradually squeezes itself, and returns back to center. |

## 7 Materiallity experiencing setup *(SYNTHESIS)*

As the selected analogue material will primarily provide the texture and materiality for the design artifact, choosing the appropriate one was key for this process. First, we advanced the materiality experiencing setup by integrating a 6DOF Stewart platform [31] with two vacuum pumps to manipulate the attached balloon. Then, we explored and examined different combinations of material and transition techniques. Exploring the defined transition samples synthesis (see p.4), we noticed the captivating capability of the pufferfish's spikes which would stand out from its skin. At rest, the spikes align with the skin forming a solid surface. However, when inflated, the spikes deviate from the skin and start to stand out, resulting in a thicker and less dense outer shell. We also sought this quality in our material samples, in which the observed density of the outer shell would change in response to the changes in shape and/or form. This quality was found in a fabric that had numerous small 'hairs' attached to it, which would stand out, resembling the movements of the spikes.

## 8 Computational mechanism explorations *(ANALYSIS)*

To have precise and comprehensive control over the materiality, we iteratively explored various computational mechanisms. We introduced pneumatic containers for customizing the shape transitions as well as capacitive sensors for human behavior detection.

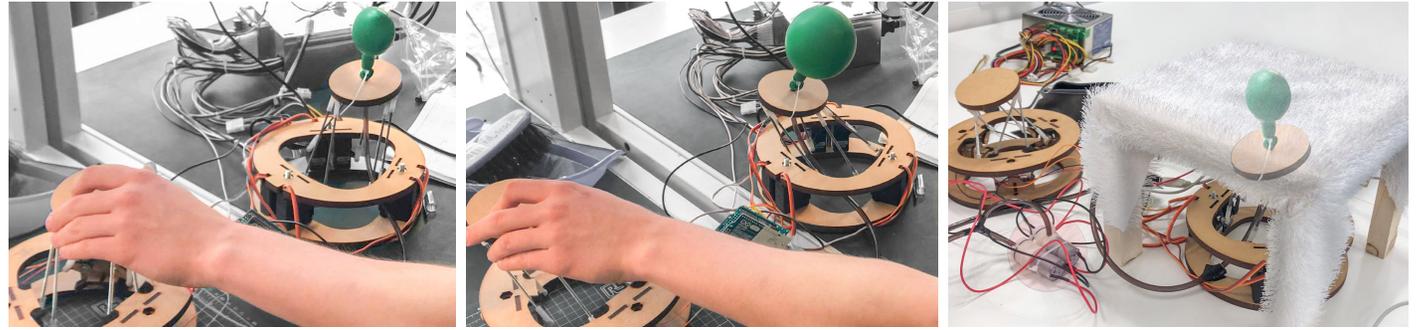

**Pneumatic container**

Inspired by PneUI[40], we implemented airtight *pneumatic containers* resembling [28,35] that can be inflated by a vacuum pump with manual control by a switch. We iteratively explored different inflatable structures and elastic materials to find the best inflation quality.

**Actuation integration**

Then, we integrated the container with the prior actuating prototype and used it to evaluate the quality of shape transitions and behavior movements with semi-manual control over a laptop.

**Sensing integration**

Next, we introduced a capacitive sensor for human behavior detection and exposed the sensing connectors with four sheets of aluminum foil distributed at the corners. This way, the prototype knows where and how close the user is and takes action spontaneously.

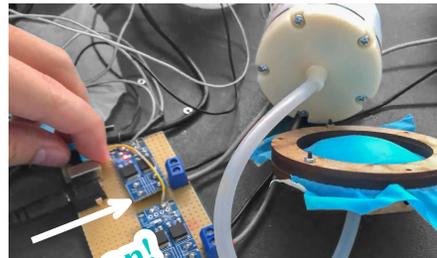
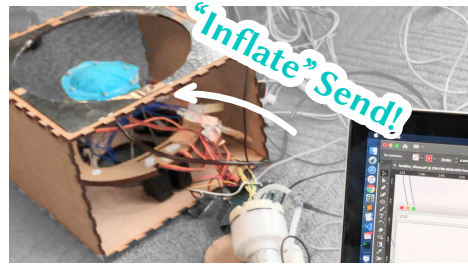
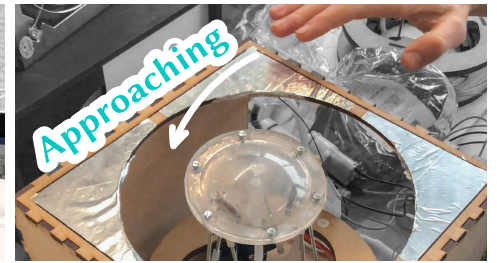
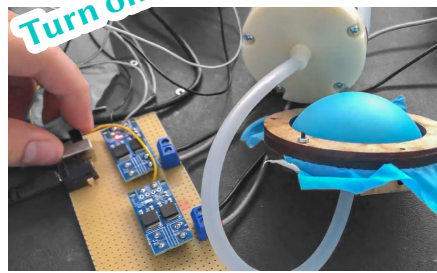
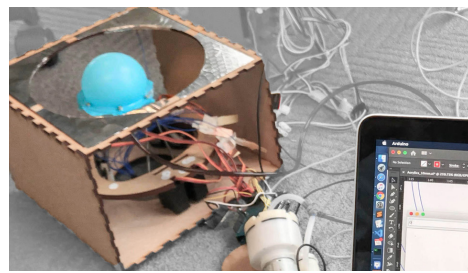
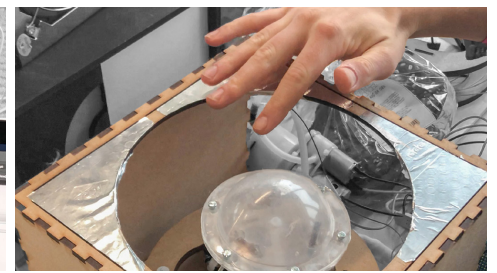

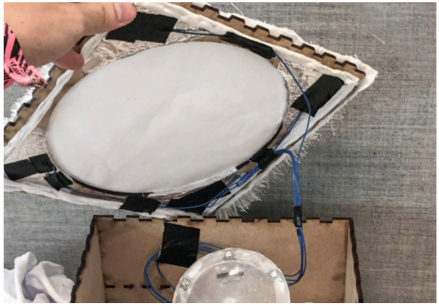
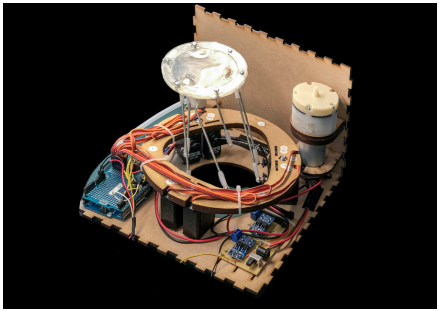
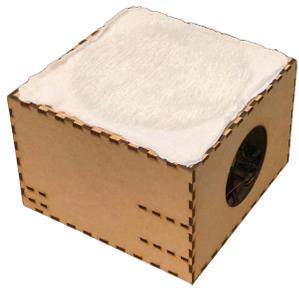

**Initial computational whole**

From top to bottom, the initial computational whole Puffy consists of a white hairy fabric that covers four sheets of aluminum foil wired with a capacitive sensor. At the center of Puffy's inside, a Stewart platform is positioned to actuate a custom inflatable that manipulates the fabric through inflation, deflation, and circular actuation.

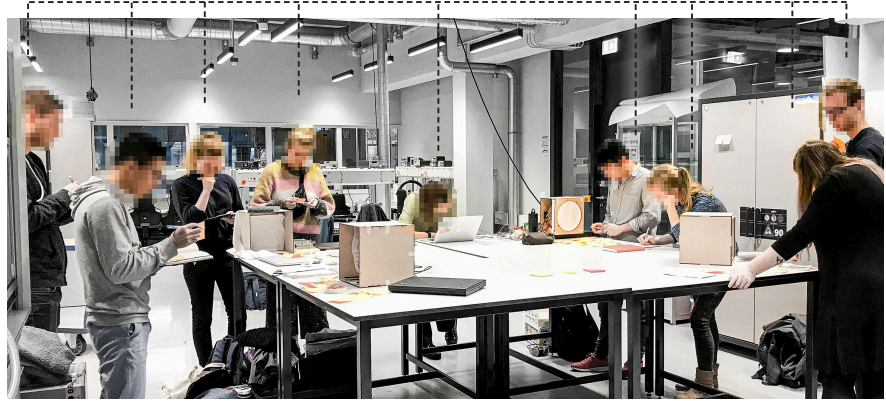

Participants (N=9) with design background at MSc. level

## 10 Design critique (ANALYSIS)

To capture user perceptions of and experiences with Puffy, the first and second authors of this pictorial hosted a design critique session. This session was aimed to help gain different perspectives of feedback for materiality experiences improvements in the consecutive stage. All recruited participants (N=9) were asked to follow the steps indicated in the section below.

### 9 Affirming the artifact

This step intends to grasp and feel the context. Participants leveraged their modality of touch and bodily movements to interact with the artifact. They experienced both visual effects and appreciated their sensational responses to the behavior of the artifact.

### Interpreting the emotion

Once their first-person experience with the artifact was gained, participants interpreted the relationship between the behavior and the emotion of the artifact they perceived while interacting with it.

### Reflecting the symbolic notions

Participants reflected on what messages the artifact and/or the designer intended to convey. These reflections were written on pink sticky notes. After that, they left questions and suggestions regarding the artifact on the green ones.

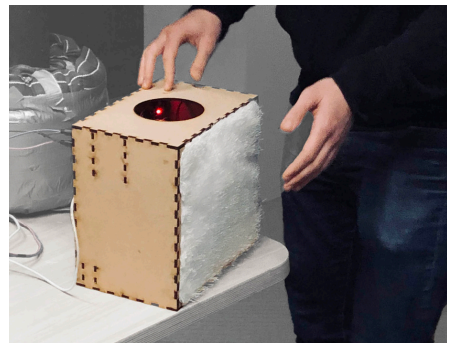
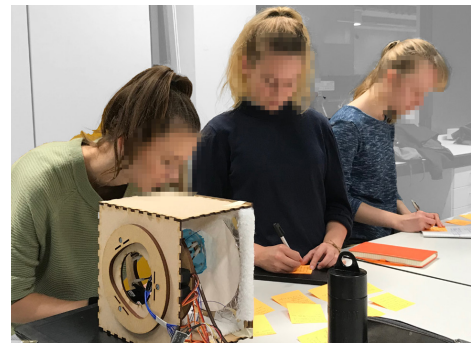
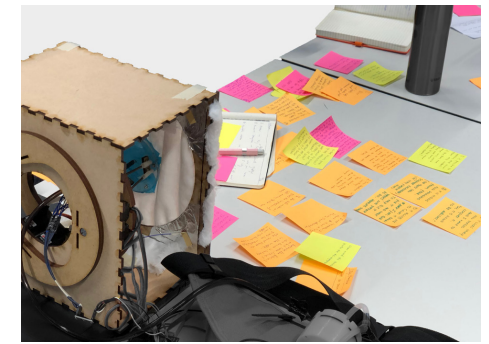

## 11 Analyzing feedback *(ANALYSIS)*

The resulting remarks and questions were analyzed through Affinity Diagramming [18] and clustered into five main categories. We learned that the fabric's visual expression and the touching of the pneumatic object underneath contradicted each other. The furry and soft property gives viewers a sense of inviting and touching. However, the sturdy pneumatic material underneath the top surface gives an opposite feeling when petting it. Most of the participants indicated that the initial breathing behavior was calm and humble, but it later became aggressive when they approached the fabric. Moreover, we learned that participants were confused about the delay of Puffy's reaction. In some cases, the fabric and actuators did not synchronize well while reacting to a user's interaction.

## 12 Synthesizing feedback *(SYNTHESIS)*

We acknowledged that the invitingness of the surface material and the aggressive behavior conflicted too much. During the final step of our process we aimed to bring these two conflicting aspects more towards each other, to further align the physical form, temporal form, and interaction gestalt [3].

# 13 Detailing the design artifact (*DETAILING*)

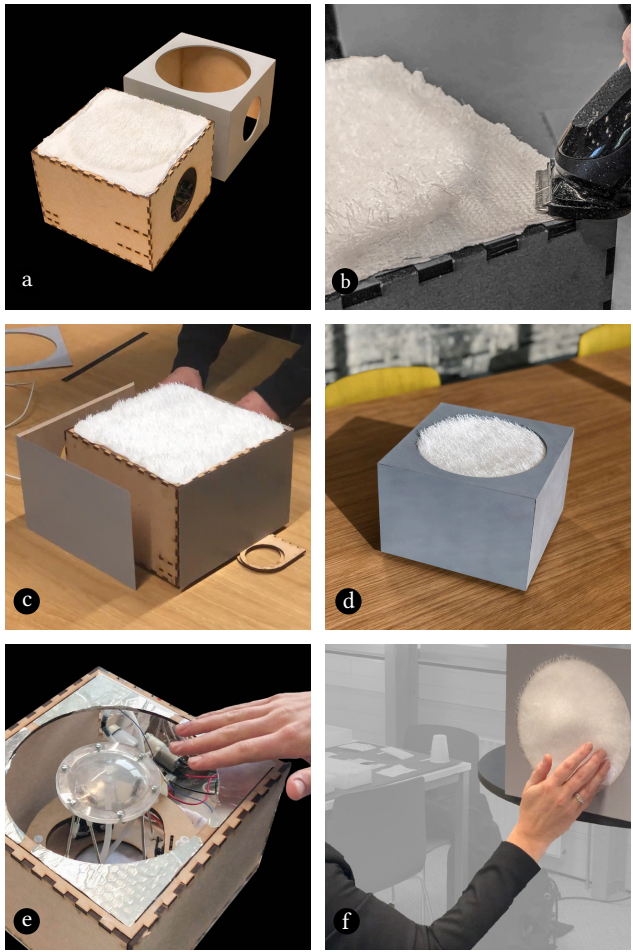

ⓐ Nested casing for seamless and aesthetic appearance
ⓑ Shaving unwanted hairy fabric for seamless assembly
ⓒ Attempting to assemble
ⓓ The final assembly
ⓔ Adjusting the actuators with slower and more fluent movements
ⓕ Vertically placing the artifact high off the ground to ensure hands approaching from the edges

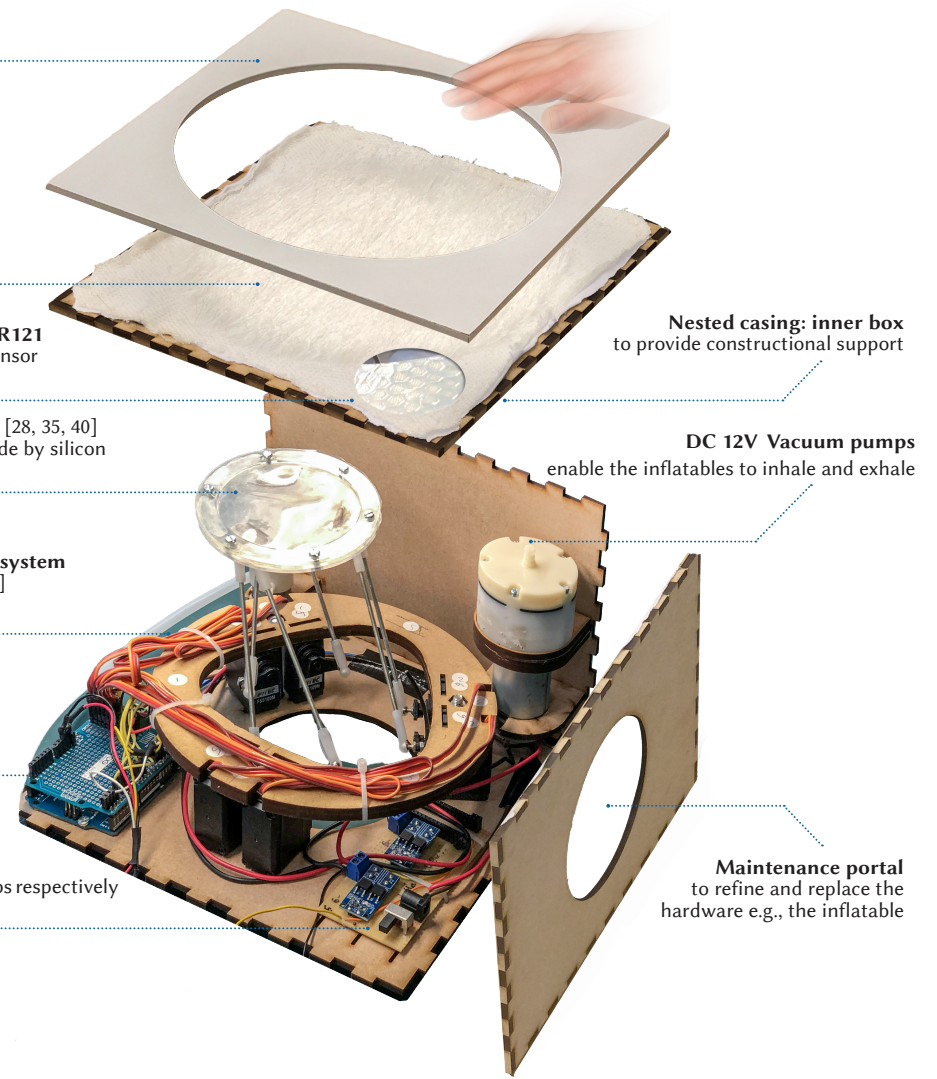

**Nested casing: outer box** to create seamless finish

**Cilia-mimetic surface**

**Aluminum foil** connected with 12-pin **MPR121** capacitive sensing touch sensor

**The inflatable** inspired by [28, 35, 40] shaped elastic material made by silicon

**Stewart 6-axis actuating system** modifications based on [31]

**Arduino UNO MCU**

**MOSFET amplifiers** to control the vacuum pumps respectively

**Nested casing: inner box** to provide constructional support

**DC 12V Vacuum pumps** enable the inflatables to inhale and exhale

**Maintenance portal** to refine and replace the hardware e.g., the inflatable

# Reflections

In this pictorial, we described the process of transforming lively inspirations into a cilia shape-changing artifact through an interactive materiality lens. Although our main approach is based on the three steps, i.e., Analysis, Synthesis, and Detailing (A-S-D) proposed in [30], from our first-person practice and instructions from Materiality of interaction [37], we found that it was not adequate to apply the approach for one round. Alternatively, the design activities (A-S-D) interlaced recursively several times, or in Wiberg's words, "*work back and forth between wholeness to details*" [37], to deepen our practical comprehension of materiality, as well as to evaluate and finetune the material experience and haptic qualities. The interlacing process is not to simply redo analysis or redesign concepts. Instead, we iteratively tackled the complexity of digital and analog materials and continually extracted insights for better material experiences.

Throughout the study, we found this design process resembles artists' creation process. Firstly, many artists have been cultivating to gain inspiration from nature (e.g., sketches of humans, animals, or plants). This consequently allows them to build a great repertoire of inspirational ideas. At the beginning of our design process, we looked for a nature analogy and were inspired by the pufferfish's form-changing and repelling behaviors. Secondly, they both do intense analysis and synthesis in the creation process. For instance, when *Pablo Picasso* created the famous painting *The Bull* in 1945, he went through several iterations to analyze the shape of a bull and portray it from hooved, horned, and muscled life-like to an abstract representation without losing identifiability. Such a process mirrors the '*Synthesis*' activity of our approach that a designer synthesizes the most valuable elements from tons of inspirations. Lastly, they both require careful attention to the subtle expressivity of the materiality and created forms. A sculptor constantly negotiates the texture and its expressivity with his hands and eyes. Similarly, in our design process, we continuously explored the subtle touch feelings with alternative materials [14] and temporal forms [36] to investigate their expressivity through the creator's first-person experiences. In addition to the mentioned similar parts, there were salient differences in terms of design considerations and evaluation. Before entering the fine-tuning step of *detailing the design artifact*, we recruited participants for a *design critique* to collect, analyze and synthesize feedback from a *third-person* point of view. However, this might not be necessary for artists during their creation of art pieces.

Based on our practice, we thought our mixed materiality approach could instigate designers to focus on the emotional or experiential aspects in addition to the pragmatic or functional features of shape-changing interfaces (SCIs) [6,21]. It may involve various creators and researchers interested in SCIs and TUIs to explore new design forms that deliver meaningful and affective experiences. By reflecting on our journey of design, we now summarize a set of design implications that are intended to inform interaction practitioners taking a similar approach:meaningful and affective experiences.

**Leverage the open-endedness and unfinishedness in the early stage.**

As described previously, in the early stage of this design approach, we started with extensive freedom for design explorations. While such a large amount of freedom also created uncertainty, in the end, we recognized the benefits of having multiple open-ended directions to explore. Such open-endedness might be a key to success in interactive materiality design. It enables us to avoid falling into the "trap" of being self-constrained by the availability, feasibility, or functionality of the materials we need in the coming phases. Also, just like other designerly or artistic crafting processes (e.g., with clay or wood), interactive materiality design heavily relies on the designer's embodied comprehension or the tacit feel for the designed materiality. And the freedom for exploration in the early stage ensures that the designers can conduct broad experimentations along various open-ended routes, which enables them to develop a sufficient feel for the crafting materials at hand. And this will extensively benefit the later design stages in which they need to decide which materials to use or how to polish the chosen materials further. And in such early open-ended experimentations, we also recommend that practitioners should not pursue the 'perfect' design samples but feel comfortable with the unfinishedness of the samples (e.g., taking a quick-and-dirty technique) so that the experimentations could yield richer design possibilities or options.

**Enable multi-modal appreciation and documentation throughout the process.**

Another implication we gained from reviewing this project is to have rich documentation that captures the subtlety, and temporality of shape-changing, for example, recording videos or sketching transition graphs as shown on p. 4. The purpose of having such documentation is not only for post-hoc analysis or inspiration for future practice, but also very much needed in supporting sensemaking and decision-making throughout the design process. In retrospect, the process started with selecting inspirations through abstracting features and fabricating and evaluating the materiality. Throughout these steps, we centralized our designer's embodied perception and comprehension of the material composites or materiality to enable our best multi-modal appreciation. As mentioned earlier, our design process has been constituted by multiple rounds of *Analysis, Synthesis, and Detailing*. In each round, we also heavily built upon the design rationales generated in previous rounds. However, much of our design experimentations, evaluations, and decisions cannot be fully communicated by texts, but we need to rely on visual communication. On the other hand, visual documentation can also help designers review how their decisions were made since design decisions can sometimes go intuitive and unconscious. Rich visual documentation using video clips, photos, or sketching could therefore benefit communication and deliberation throughout the process.

**Emphasize the hedonic and experiential aspects in the exploration.**

As demonstrated in our process, our exploration has focused heavily on the nuances of the hedonic and experiential aspects of the materiality and the designed artifact. And we recognize this as an advantage of such interactive materiality approach, which could complement the design approaches that purely focus on the pragmatic aspects of design (e.g., utility, usability, or efficiency). With our addressed approach, much of the designer's attention could be effectively directed to the subtle differences of the sensory, experiential, and aesthetic aspects of the interactive artifacts, with the very depth that is often not likely to achieve in pragmatic approaches of design. From our own experiences, such an approach could shift designers from a problem-solving mindset to a curiosity-driven mindset, and help them get immersed in the playful, embodied, and purposeless experimentations with computational and analog materials at hand. Therefore, we recommend that such an interactive materiality approach could be more widely adopted as a complementary or additional method to traditional interaction design processes so that the designers could be equally facilitated in both exploring the pragmatic qualities and the hedonic qualities.

In short, we provide the following takeaways for those intending to leverage bio-inspiration for interactive materiality:

1. during the early stage, try *freely* exploring inspiration sources; try *appreciating* them and *documenting* the shape transitions, material textures, and underlying intentions in a *multi-modal* manner; and avoid concerning the technical and engineering feasibility;
2. while mimicking, try *reasoning* the relationship between the selected natural element and its intentions/surroundings; try mimicking that with *rough* setups to experience effects addressed on potential viewers;
3. while fabricating, try hacking, mixing, and iterating prototypes with parts from *daily life* or the *HCI community*;
4. lastly, try *centralizing sensations* and *experiences* throughout all steps, aiming to reveal *nuanced* qualities rather than exactly replicating the natural elements.

**Future work**

When considering potential application areas, stimulating multi-modal and haptic-rich user-product relationships seem prominent. By exploiting the physically rich characteristics of interactive materiality instead of using vocal or graphical UIs, we might be able to achieve the goal of creating physically rich interactions [8,9]. Specifically, as Puffy reveals different degrees of shape changes, it might benefit some fabric-wrapped yet screenless devices such as internet-of-things virtual assistants (e.g., Apple HomePod, Google Home mini) or smart home appliances such as OSKARRR [32] to exhibit rich haptic experiences as well as anthropomorphic attributes with dynamic forms. In these ways, Puffy's cilia-mimetic surface could change its shape in accordance with the mood or intention of the interactive systems.

**Limitations**

Although we used the puffer fish which is more object-like as inspiration, the end artifact turned out to be more surface-like. This might be self-limiting in terms of interaction possibilities. However, as we aimed to focus solely on the material's quality and the designed interactions, for this project, we argued for experiencing and evaluating materiality in an exploratory and first-hand manner over biologically or mechanically mimicking a puffer fish. With these considerations in mind, we leaned towards a quick and flexible setup that allowed us to swap material samples handily and promptly. Otherwise, we might fall into the 'trap' of concerning technical and engineering matters.

# Suggestions for prototyping

**High degrees of durability with high degrees of hot-swap functionality**

Shape-changing prototypes need to be highly durable as they are expected to interact physically. Yet, the materiality experience only works the best and most authentic when all materials and elements are entangled as an assembled whole. That means, the materiality shall be experienced and evaluated through the computational whole instead of any division part of it. However, as modifications might be applied during the fine-tuning process, it might be tricky if a maintenance portal or so-called the 'backdoor' is not configurated beforehand. Hereby, we highlight the hot swap functionality design in designing interactive materiality.

**Hot-swap 1: maintenance portal - the "backdoor" design**

Experiencing and evaluating interactive materiality requires all materials, both digital and analogue ones, to be integrated as a whole. And yet, during our materiality experience examination with the initial computational whole, the inflatables blew out serval times due to pump overinflation caused by false sensing detection. The disassembling and reassembling of the MDF box made us to implement a "*backdoor*" for replacing the inflatables handily. The portal also benefited us when fine-tuning Puffy with different thicknesses of rubber for the inflation. Thus, while crafting, question yourself:

1. Will your prototype involve *consumables* that are fragile?
2. Would it be convenient to *replace* them when they are assembled inside a computational whole, and how?

**Hot-swap 2: nested casing**

To ensure both durability and hot-swap functionality, we suggest building the computational whole in a nested structure. For example, Puffy is configured within a double-nested box: the inside one (assembled as a tooth-and-slot box) provides construction, and the outside one (painted in grey) creates a seamless appearance of Puffy.

**Hacking and mixing materials coming from daily life or the HCI community**

**iterations**
process inspired by [28,35,40]

**New compositions or customized materials**

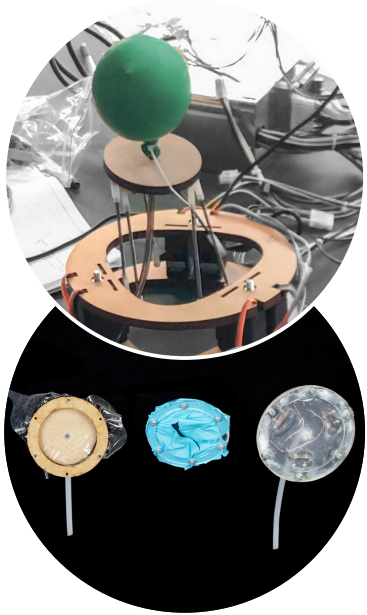

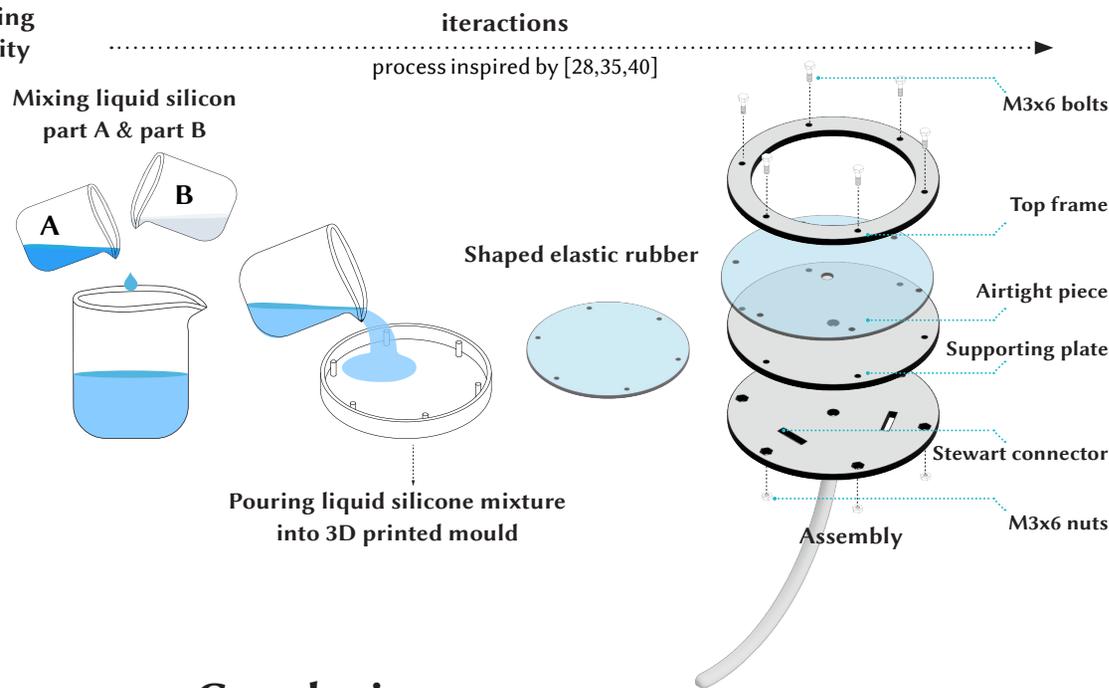

**Mixing liquid silicon part A & part B**

A  B

**Shaped elastic rubber**

**Pouring liquid silicone mixture into 3D printed mould**

**Assembly**

M3x6 bolts
Top frame
Airtight piece
Supporting plate
Stewart connector
M3x6 nuts

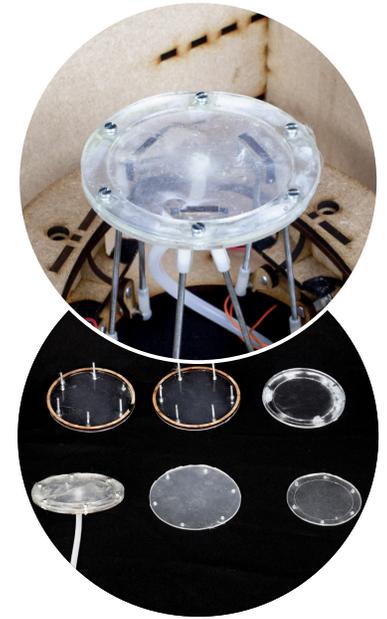

### Hacking, mixing, and iterating prototypes as an approach

Before originating your prototype for creating materiality, we recommend hacking, mixing the existing prototypes from everyday life or the HCI communities, and trying to iterate those prototypes into a satisfying form. In reviewing the process of crafting Puffy, we attempted to hack, e.g., Ink printer or Joystick Stewart considering their actuating systems; we tried to integrate the inflatables with the *Stewart 6DOF platform*, which both have been extensively investigated in the field of morphing material and robotics respectively and incorporated sensing technologies and fabric materials iteratively to create the desired pufferfish-alike materiality experiences. While iterating, we started with the most accessible materials (e.g., balloons, latex gloves, etc.) to familiarize ourselves with the physicality and gradually shifted to the exploration and fabrication of custom-shaped materials.

## Conclusion

As material, sensors, and actuators are becoming more entangled in forming new interactions, HCI scholars have started to take materiality as an entry point for conducting design research. Nonetheless, design cases and instructions that show designerly ways of attending materiality in design practices are still needed. This pictorial presents a concrete case of designing a shape-changing artifact using the materiality approach. The approach consists of three key activities (analysis, synthesis, and detailing) interlaced recursively along the whole design process. The 'analysis' activity gained nature-inspired analogy and iteratively explored from shape transitions through analog materials and computational mechanisms to gain an understanding of the design context; The 'synthesis' activity synthesized findings regarding digital and analog materials, both self-reflection and peers' critiques, to navigate the consecutive activity; The 'detailing' activity encoded the designers' symbolic notions of the interactive materiality as well as the synthesized critique from the audience into a set of iterative prototypes. Our reflection surfaced the value of having such an interlaced iterative process. As a result, by offering a reflective analysis of our approach, we contribute an illustrative step-by-step guide of our highly embodied design process and a set of practical implications and suggestions, to inspire future creators to design interactions with interactive materiality.

### Acknowledgment

The authors want to appreciate the technical support from Jasper and Chet at the Rapid Prototyping lab (TU Eindhoven), Material Workshop arrangement by Simone G. de Waart, substantial suggestions from the anonymous reviewers, kind advice from Prof Stephan Wensveen and Jeffrey Ho, supplementary video clip by Wei Lai. This project is supported by the UGC Funding Scheme (RHCE & G.73.xx.R006) from The Hong Kong Polytechnic University.